\documentclass[twocolumn,showpacs,preprintnumbers]{revtex4}
\usepackage{amsmath}
\usepackage{dcolumn}
\usepackage{bm}
\usepackage{graphicx}
\usepackage{amsfonts}
\usepackage{amssymb}
\usepackage{mathrsfs}
\usepackage{color}
\usepackage{braket}

\begin{document}

\title{Parity breaking and scaling behavior in the spin-boson model}
\author{Tao Liu$^{1,2}$}
\email{liutao849@163.com}
\author{Mang Feng$^{2}$}
\email{mangfeng@wipm.ac.cn}
\author{Lei Li$^{1}$}
\author{Wanli Yang$^{2}$}
\author{Kelin Wang$^{3}$}
\affiliation{$^{1}$ The School of Science, Southwest University of Science and
Technology, Mianyang 621010, China \\
$^{2}$ State Key Laboratory of Magnetic Resonance and Atomic and
Molecular Physics and Key Laboratory of Atomic Frequency Standards,
Wuhan Institute of Physics and Mathematics, Chinese Academy of
Sciences, and Wuhan National
Laboratory for Optoelectronics, Wuhan, 430071, China \\
$^{3}$ The Department of Modern Physics, University of Science and
Technology of China, Hefei 230026, China}

\begin{abstract}
We study the breaking of parity in the spin-boson model and
demonstrate unique scaling behavior of the magnetization and
entanglement around the critical points for the parity breaking
after suppressing the infrared divergence existing inherently in the
spectral functions for Ohmic and sub-Ohmic dissipations. Our
treatment is basically analytical and of generality for all types of
the bath. We argue that the conventionally employed spectral function is
not fully reasonable and the previous justification of quantum phase
transition for localization needs to be more seriously reexamined.
\end{abstract}

\maketitle

The spin-boson model (SBM) has been key to phenomenological
descriptions of open quantum systems, in which the environment acts
as a bosonic bath responsible for dissipation of the system, i.e.,
the spin \cite{weiss, leggett}. Besides the coherence of the spin, the
correlation between the spin and the bath degrees of
freedom has also attracted much attention.

The SBM hamiltonian is given by \cite{leggett}
\begin{equation}
H=\frac {\epsilon}{2}\sigma_{z}-\frac {\Delta}{2}\sigma_{x}+
\sum_{k}\omega_{k}a^{\dagger}_{k}a_{k} + \sum_{k}\lambda_{k}(a^{\dagger}_{k}
+ a_{k})\sigma_{z},
\end{equation}
where $\sigma_{z}$ and $\sigma_{x}$ are usual Pauli operators,
$\epsilon$ and $\Delta$ are, respectively, the local field
(also called c-number bias \cite {leggett}) and the
tunneling regarding the two levels of the spin. $a^{\dagger}_{k}$
and $a_{k}$ are creation and annihilation operators of the bath
modes with frequencies $\omega_{k}$, and $\lambda_{k}$ is the
coupling between the spin and the bath modes, which is governed by
the spectral function $J(\omega)=\pi\sum_{k}\lambda_{k}^{2}\delta
(\omega-\omega_{k})$ for $0<\omega <\omega_{c}$ with the cutoff
energy $\omega_{c}$. In the infrared limit, i.e.,
$\omega\rightarrow$0, the power laws regarding $J(\omega)$ are of
particular importance. Considering the low-energy details of the
spectrum, we have $J(\omega)=2\pi\alpha\omega_{c}^{1-s}\omega^{s}$ with
$0<\omega <\omega_{c}$ and  the dissipation strength $\alpha$. The
exponent $s$ is responsible for different bath with super-Ohmic bath
$s>$1, Ohmic bath $s =$1 and sub-Ohmic bath $s <$1.

There are several approaches solving the SBM, such as the
non-interacting blip approximation \cite {leggett}, numerical
renormalization group (NRG) \cite
{vojta2,vojta1,hur,Bu1,and,rmp1,Bu2,Bu3,cheng}, quantum Monte Carlo
(QMC) \cite{winter} and so on \cite {AA,chin}. The main concern in the SBM
is for the localization of the spin, and quantum phase
transition (QPT) between the delocalization and localization in the
Ohmic dissipation has been well investigated so far \cite
{hur-review}. But the second-order QPT with sub-Ohmic dissipation,
which is currently under intensive investigation, is not yet fully
understood.

In contrast to the intensively studied QPT, we investigate the breaking of parity in the
SBM. We show that variation of the parameters in Eq. (1) leads to
different symmetries of the SBM hamiltonian and the parities to be broken
are responsible, respectively, for
localization and delocalization. The key step in our treatment is
the suppression of the infrared divergence existing in the spectral
functions for Ohmic and sub-Ohmic dissipations, which enables us to
demonstrate unique scaling behavior of the magnetization and
entanglement in the vicinity of critical points for the parity
breaking. More importantly, our treatment is basically analytical
and suitable for all types of the bath, by which we can fully
understand the physics behind the infrared divergence and the
scaling behavior.

We start from suppression of the infrared divergence
in the spectral functions. With reference to the standard form of
the spectral function $J(\omega)$, we introduce a distribution
function $\rho(\omega)=1-e^{-P(\omega/\omega_{c})^{2}}$ with $P$ a
very large number, which is a smooth variation in the function of
$\omega$ with $\rho(0)=0$. So the spectral function is modified as
\begin{equation}
J'(\omega)=\pi\sum_{k}\lambda_{k}^{2}\delta(\omega-\omega_{k})\rho(\omega)=2\pi\alpha\omega_{c}^{1-s}\omega^{s}\rho(\omega),
\end{equation}
which fits $J(\omega)$ very well, as shown in Fig. 1 in the case of
$P=10^{6}$. However, using $J'(\omega)$, the values of integration
near the zero frequency could be effectively suppressed due to the
exponential factor in $\rho(\omega)$.
\begin{figure}[htbp]
\label{Fig1}
\includegraphics[width=2.6in]{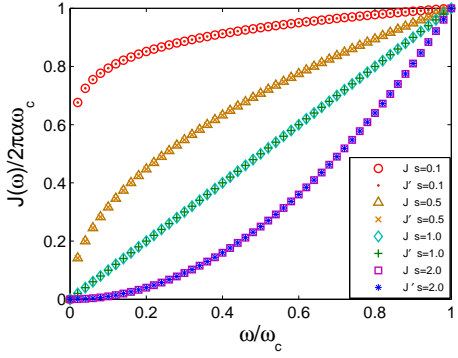}
\caption{(color online) Comparison between $J'(\omega)$ and $J(\omega)$ with $P=10^{6}$.
$J'(\omega)$ fits $J(\omega)$ very well with the difference hard to be distinguished from the curves.}
\end{figure}

To check how well the modified spectral function works, we compare
calculations in the following with $J(\omega)$ and $J'(\omega)$. We
solve the SBM using displaced coherent states \cite {TLiu,Zhang} as
the eigenfunction of Eq. (1), i.e.,
$$|\Psi\rangle=
\begin{pmatrix}
\sum_{\{n\}} c_{\{n\}} |\{n\}\rangle_{A} \\
\sum_{\{n\}} (-1)^{\sum_{k}n_{k}+1}d_{\{n\}}|\{n\}\rangle_{B}
\end{pmatrix},$$
where $c_{\{n\}}$ and $d_{\{n\}}$ are coefficients to be determined
later and $\{n\}=n_{1}, \cdots, n_{N}$ are for different bosonic
modes. $|\{n\}\rangle_{A(B)}$ is the product of displaced coherent
states of different modes, i.e.,
$|\{n\}\rangle_{A(B)}=\prod_{k=1}^{N}|n_{k}\rangle_{A_{k}(B_{k})}$,
where
\begin{eqnarray}
|n_{k}\rangle_{A_{k}} =\frac {e^{-q_{k}^{2}/2}}{\sqrt{n_{k}!}}(a^{\dagger}_{k}+q_{k})^{n_{k}}e^{-q_{k}a_{k}^{\dagger}}|0\rangle, \notag\\
|n_{k}\rangle_{B_{k}} =\frac {e^{-q_{k}^{2}/2}}{\sqrt{n_{k}!}}(a^{\dagger}_{k}-q_{k})^{n_{k}}e^{q_{k}a_{k}^{\dagger}}|0\rangle, \notag
\end{eqnarray}
with the displacement variables $q_{k}=\lambda_{k}/\omega_{k}$ and
$k=1, 2, \cdots, N$. Using Schr{\"o}dinger equation, we have, in the case of $\Delta\ll 1$,
\begin{equation}
\Big[\sum_{k}\omega _{k}(m_{k}-q_{k}^{2})+\frac {\epsilon}{2}\Big]c_{\{m\}} + \frac{\Delta }{2}
d_{\{m\}}D_{\{m,m\}} =Ec_{\{m\}},
\end{equation}
\begin{equation}
\Big[\sum_{k}\omega _{k}(m_{k}-q_{k}^{2})-\frac {\epsilon}{2}\Big]d_{\{m\}} + \frac{\Delta }{2}
c_{\{m\}}D_{\{m,m\}} =Ed_{\{m\}},
\end{equation}
where other terms, except $D_{\{m,m\}}$, in $D_{\{m,n\}}$ have been
neglected due to the reasons in Supplementary Material \cite {SM1}.
$D_{\{m,m\}}$ is given by \cite {TLiu,Zhang}
$$e^{-2\sum_{k}q_{k}^{2}}\prod_{k=1}^{N}\sum_{j=0}^{m_k}(-1)^{j}\frac{m_{k}!(2q_{k})^{2m_{k}-2j}}{[(m_{k}-j)!]^{2}j!}.$$
It is straightforward to yield following solutions from Eqs. (3) and
(4),  that is, the eigenenergies
$E_{\{m\}}^{\pm}=\sum_{k}\omega_{k}(m_{k}-q_{k}^{2})\pm\sqrt{\epsilon^{2}+\Delta^{2}D^{2}_{\{m,m\}}}/2,$
and the coefficients
$c^{\pm}_{\{m\}}=\mu^{\pm}_{\{m\}}/\sqrt{1+(\mu^{\pm}_{\{m\}})^{2}}$
and $d^{\pm}_{\{m\}}=1/\sqrt{1+(\mu^{\pm}_{\{m\}})^{2}}$ with
$\mu^{\pm}_{\{m\}}=\Big[\epsilon\pm\sqrt{\epsilon^{2}+\Delta^{2}D^{2}_{\{m,m\}}}\Big]/\Delta
D_{\{m,m\}}$. It is obvious from the expression of eigenenergies
that the ground-state energy $E^-_{\{0\}}$ is smaller than
$E^+_{\{0\}}$.

The above analytical treatments for Eq. (1) can be considered as
complete and reliable solutions for the characteristic of the SBM \cite {SM2}.
For our purpose, we may
focus on the ground-state characteristic of the model to see
counter-intuitive phenomena in the SBM. In such a case, the infrared
divergence is reflected in the variable $D_{\{0,0\}}$, which is
written as $D_{\{0,0\}}=e^{-2\sum_{k}q_{k}^{2}}$ \cite {explain0}.
We first check $q_{k}^{2}$ in the case of the bath modes of the
continuous spectrum \cite{vojta2,vojta1}. From the conventional spectral
function $J(\omega)$, we have
\begin{eqnarray}
\sum_{k}q_{k}^{2}=\sum_{k}\lambda_{k}^{2}/\omega^{2}_{k} =\int_{0}^{\omega_{c}}
\sum_{k}\frac {\lambda_{k}^{2}}{\omega^{2}}\delta(\omega-\omega_{k})d\omega  \notag \\
=\int_{0}^{\omega_{c}}2\alpha\omega_{c}^{1-s}\omega^{s-2}d\omega =
2\alpha\beta, ~~~~~~~~~~~~~
\end{eqnarray}
with
\begin{displaymath}
\beta=\left\{ \begin{array}{ll} \frac{1}{s-1} [1-(\frac{\omega_c}{\omega_1})^{1-s}] & \textrm{if $s<1$} \\
\ln (\frac{\omega _{c}}{\omega_{1}})  & \textrm{if $s=1$} \\
\frac{1}{s-1} & \textrm{if $s>1$}, \end{array} \right.
\end{displaymath}
where $\alpha$ and $\omega_{c}$ are defined above in the spectral
function. $\omega_{1}$ is a small quantity regarding the frequency
difference from $\omega=0$. In the case of the infrared limit, i.e.,
$\omega_{1}\rightarrow 0$, we have
$\sum_{k}q_{k}^{2}\rightarrow\infty$ if $s\leqslant 1$, which is
actually caused by the uncertainty in the spectral function for
$\omega_{1}\rightarrow$0.  However, if using the modified spectral
function $J'(\omega)$ and repeating Eq. (5), we have
$\sum_{k}q_{k}^{2}=2\alpha\beta'$ with
\begin{displaymath}
\beta'=\left\{ \begin{array}{ll} \frac{1}{s-1}+ \frac {1}{2P^{(s-1)/2}}[\Gamma(\frac {s-1}{2}, P)
-\Gamma(\frac {s-1}{2})] & \textrm{if $s\neq 1$} \\
\frac{1}{2}[\gamma + \ln(P) +\Gamma(0,
P)] & \textrm{if $s=1$},
\end{array} \right.
\end{displaymath}
where $\gamma$ is the Euler-Mascheroni constant, $\Gamma(\cdot)$ is
the gamma function and $\Gamma(\cdot,\cdot)$ is the upper incomplete
gamma function. Since the gamma functions are finite even for a very
large value of $P$, $\beta'$ is definitely convergent. Similar
results could be obtained for the bath modes of the discretized
spectrum \cite {SM3}.

We have noted the results in previous publications that there are
QPTs in Ohmic and sub-Ohmic dissipation cases, but not in
super-Ohmic one, which exactly corresponds to the infrared
divergence demonstrated above: Divergence for Ohmic and sub-Ohmic
bath, but convergence for super-Ohmic bath. As a result, it is
reasonable to presume that the QPT presented previously are probably
induced totally or partially by the infrared divergence in the
calculations using $J(\omega)$. In fact, there have been some
discussions about the shortcomings in NRG methods, which cause the
qualitatively incorrect results when studying quantum-critical
phenomena, and spoil the determination of critical exponents and
behaviors \cite{Bu2}. Additionally, there were hints that the NRG
displays truncation errors or other errors in the long-range ordered
phase \cite{Bu2,Bu3,Tong,arxiv2011}.

To fully understand the characteristic of the SBM, we may first
consider two special cases of Eq. (1), where we denote the case of
$\epsilon\neq 0$ with $\Delta=0$ ($\epsilon=0$ with $\Delta\neq0$) by
$H(\epsilon\neq 0, \Delta \!=\! 0)$ ($H(\epsilon =0, \Delta\neq
0)$).  For the two special hamiltonians, we introduce, respectively,
two parity operators $\Pi_I=\sigma_z$ and $\Pi_{II}=\sigma_x e^{i
\pi \sum_k a^\dagger _k a_k}$. For $H(\epsilon\neq 0, \Delta= 0)$, we have $[H(\epsilon\neq 0, \Delta= 0), \Pi_I]=0$,
with the ground state of their common eigenfunction to be $\ket{\psi_{I,0}^-} =
\begin{pmatrix}
0 \\
\ket{\{ 0 \}}_B
\end{pmatrix}$
satisfying $\Pi_I\ket{\psi_{I,0}^-}=-\ket{\psi_{I,0}^-}$, i.e., an odd parity state of $\Pi_I$. Since
$\bra{\psi_{I,0}^-}\sigma_z \ket{\psi_{I,0}^-} =-1$,
$\ket{\psi_{I,0}^-}$ is always a localized state. Similarly, we have $[H(\epsilon=0, \Delta\neq 0), \Pi_{II}]=0$. The ground
state of their common eigenfunction $\ket{\psi_{II,0}^-}
= \frac{-1}{\sqrt{2}} \begin{pmatrix}
\ket{\{0\}}_A \\
\ket{\{0\}}_B
\end{pmatrix}$
is an even parity state of $\Pi_{II}$ with $\Pi_{II}
\ket{\psi_{II,0}^-} = \ket{\psi_{II,0}^-}$.  We have
$\bra{\psi_{II,0}^-} \sigma_z \ket{\psi_{II,0}^-} = 0$, meaning
$\ket{\psi_{II,0}^-}$ to be always a delocalized state. It is evident that
the odd (even) parity breaks in the variation from $\Delta=0$ with $\epsilon\neq 0$ ($\epsilon=0$ with $\Delta\neq 0$)
to both $\Delta\neq 0$ and $\epsilon\neq 0$ because the hamiltonian $H$ in Eq. (1) never commutes with any of
the parity operators above. This also means that the ground state of $H$ would never stay forever in delocalization or localization, but possibly moving between delocalization
and localization in variation of certain characteristic parameters, such
as $\alpha$. We show below the behavior of magnetization in the
vicinity of critical points of the parity breaking.

The magnetization of the SBM is of importance to symbolize the transitions
between delocalization and localization of the model. Using the modified spectral function and the
displaced coherent states $|\{n\}\rangle_{A(B)}$, we have
\begin{equation}
\langle\sigma_{z}\rangle=(c^{-}_{\{0\}})^{2}-(d^{-}_{\{0\}})^{2}=\frac {-\kappa}{\sqrt{\kappa^{2}+e^{-8\alpha\beta'}}},
\end{equation}
where the average is made by the ground-state of the model and
$\kappa = \epsilon/\Delta$. Since $\beta'$ is convergent for any
type of the bath, $\langle\sigma_{z}\rangle$ should be of finite
value. Performing the second derivative of
$\langle\sigma_{z}\rangle$ with respect to $\alpha$, we obtain a
reflection point $\alpha_{c}=-\ln(2\kappa^{2})/8\beta'$, by which
Eq. (6) is rewritten as
\begin{equation}
\langle\sigma_{z}\rangle=\frac
{-\kappa}{\sqrt{\kappa^{2}+e^{\alpha'\ln(2\kappa^{2})}}},
\end{equation}
under the scaling transformation $\alpha'=\alpha/\alpha_{c}$. For a
fixed value of $\kappa$, the magnetization in Eq. (7) is only
relevant to $\alpha'$, rather than other characteristic parameters.
So $\alpha_{c}$ can be regarded as a scale of the dissipation
strength. In addition, if we set $\alpha'=1$, the magnetization turns to be a
constant $-1/\sqrt{3}$, which implies a fixed crossing point for
different types of the bath in the magnetization with variation of
$\alpha'$ (See Fig. 2(a,b)).
\begin{figure}[htbp]
\label{Fig3}
\includegraphics[width=2.8in]{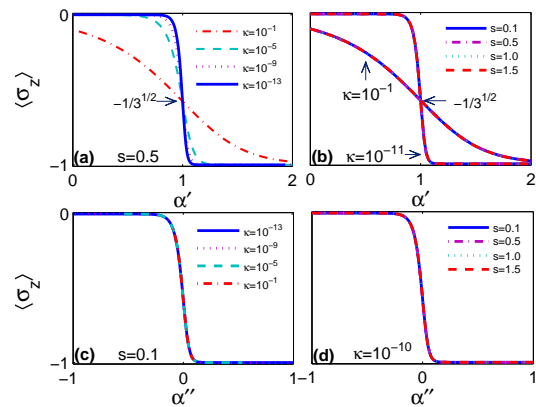}
\caption{(color online) The scaling of the magnetization, with (a):
as a function of $\alpha'$ under sub-Ohmic dissipation for different
$\kappa$; (b): as a function of $\alpha'$ for different $\kappa$ and
types of the bath; (c) and (d): as a function of $\alpha''$, which
remain unchanged for the characteristic parameters $\kappa$ and $s$
in the model.}
\end{figure}

It is more interesting to demonstrate the scaling behavior of the
magnetization with a displaced dissipation strength
$\alpha''=(\alpha-\alpha_{c})\beta'/\sqrt{27}$. Since
$\langle\sigma_{z}(\alpha'')\rangle=
-1/\sqrt{1+2e^{-24\sqrt{3}\alpha''}},$ which is independent of both
$\kappa$ and $s$ under the scaling transformation, the magnetization
with respect to $\alpha''$, as presented in Fig. 2(c,d), remains
unchanged for different types of the bath and different tunneling
and localization parameters. The scaling transformation was usually
used to find QPT around the critical points, where the scale
invariance appears in the neighborhood of the critical points. In
contrast, our results present the scale invariance in the whole
region of $\alpha''$, which can be understood as the critical behavior
resulting from the parity breaking regarding $\Pi_I$ and $\Pi_{II}$.
It could be more clarified if we check the linear variation near the region of
$\alpha''=0$ with the slope
$d\langle\sigma_{z}\rangle/d\alpha''|_{\alpha''\rightarrow 0}=-8$
(See Fig. 3), which is a  continuous change between the
delocalization and the localization without any cusp-like behavior.

It was indicated in previous studies for the Ohmic damping at $\epsilon=0$ that
quantum Kosterlitz-Thouless transition separates the delocalized
phase at small $\alpha$ from the localized phase at large $\alpha$
\cite{leggett,hur-review}. In contrast, the situation of $\epsilon=0$ in our case only
corresponds to delocalization and there is no possibility for any QPT.
However, for Eq. (1), there is possibility of translation (with no cusp-like behavior) between the
localization and delocalization in our results, where the delocalization and
localization correspond, respectively, to small $\alpha$ ($\alpha <\alpha_{c}$) and large
$\alpha$ ($\alpha >\alpha_{c}$). Nevertheless, our results is only relevant to the critical
behavior of the parity breaking and hold for not only the sub-Ohmic damping but
also other types of the bath.

\begin{figure}[htbp]
\label{Fig4}
\includegraphics[width=2.5in]{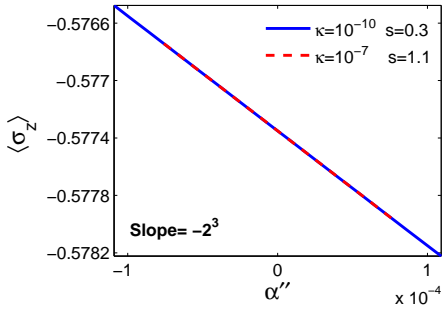}
\caption{(color online) The magnetization in variance with
$\alpha''$ in the nearby region of $\alpha''=0$ under different
characteristic parameters.}
\end{figure}

\begin{figure}[htbp]
\label{Fig5}
\includegraphics[width=2.8in]{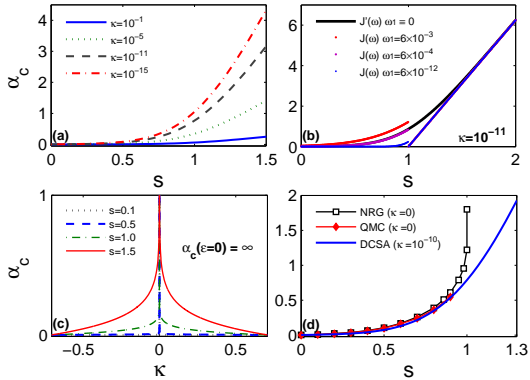}
\caption{(color online) The scale $\alpha_{c}$, where (a) as a
function of $s$ under different characteristic parameters shows the
smooth variance for all types of the bath; (b) as a function of $s$
compares conventional spectral function with the modified one for
$\kappa=10^{-11}$; (c) in variation with $\kappa$ for different
types of the bath shows the singularity at $\epsilon=0$; (d) as a
function of $s$ compare our displaced coherent-state approach (DCSA)
for $\kappa=10^{-10}$ with NRG and QMC for $\kappa=0$.}
\end{figure}

The scale $\alpha_{c}$ has some unique features: It is universal for
different types of the bath, which means a continuous and smooth
curve with respect to $s$ (See Fig. 4(a)).  In contrast, if we
employ the conventional spectral function, $\beta$ in the expression of
$\alpha_{c}$ would yield $\alpha_{c}=0$ in the case of $s\le$1, but
finite values for $s>1$, which causes drastic changes in the
variation of $\alpha_{c}$ with respect to $s$  and corresponds to
the appearance of QPT around the point $s=1$ (See Fig. 4(b)). This
is another evidence that the QPT for the localization in the SBM is
related to the infrared divergence. On the other hand, $\epsilon=0$
is a singularity in $\alpha_{c}$, as shown in Fig. 4(c), which is,
as mentioned above, due to the parity breaking regarding $\Pi_{II}$.
In this sense, any characteristic parameter calculated under
$\epsilon=0$ and $\epsilon\neq$0 should be very different. So the
fitting in Fig. 4(d) for our approach using a negligibly small
$\kappa$ with respect to the NRG and QMC at $\kappa=0$ gives the
quantitative evidence that both the NRG and the QMC suffer from the
infrared divergence for $s<$1 with the uncertainty equivalent to
the effect of $\kappa=10^{-10}$ in the calculation without the
infrared divergence.


The feature of the scaling can also be reflected in entanglement. We
denote the entanglement by von Neumann entropy
$E=-p_{+}log_{2}p_{+}-p_{-}log_{2}p_{-}$ \cite {nc} with
\begin{equation}
p_{\pm}=\frac {1}{2}\Big(1\pm\sqrt{\langle\sigma_{z}\rangle^{2}
+\langle\sigma_{x}\rangle^{2}}\Big)=\Big(1\pm \sqrt{\frac {\kappa^{2}
+e^{-16\alpha\beta'}}{\kappa^{2}+e^{-8\alpha\beta'}}}\Big)/2,
\end{equation}
which reflects the bipartite quantum
correlation between the spin and the bath. We plot in Fig. 5 the
entanglement with respect to $\alpha''$ for different $\kappa$,
where $\alpha\beta'$ in Eq. (8) is replaced by
$\sqrt{27}\alpha''-\ln(2\kappa^{2})/8$ under the scaling
transformation. Since the magnetization reaches 0
for $\alpha < \alpha_{c}$, i.e., the delocalization, and drops to -1
if $\alpha > \alpha_{c}$, i.e., the localization (Refer to Fig.
2(c,d)), we could know from Fig. 5 that the delocalization and
localization in the SBM correspond, respectively, to the increasing
entanglement and the decreasing entanglement. In this sense, the
scale $\alpha_{c}$ is also the reflection point for the entanglement
increasing and decreasing. As a result, we may easily conclude that
the ground-state of $H(\epsilon =0, \Delta\neq 0)$, which is always
in delocalization, owns the entanglement increasing due to
$\alpha<\alpha_{c}$ with $\alpha_{c}\rightarrow\infty$, and the
ground-state of $H(\epsilon\neq 0, \Delta=0)$ is always localized
with the entanglement decreasing and in most cases with
disentanglement \cite {explain2}.
\begin{figure}[htbp]
\label{Fig6}
\includegraphics[width=2.6in]{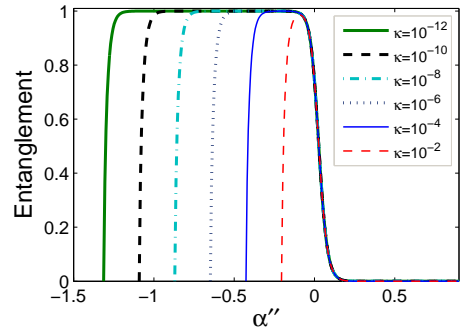}
\caption{(color online) Entanglement in variance with $\alpha''$ for
different $\kappa$. The curves remain unchanged for any value of $s$.}
\end{figure}

In comparison with \cite {chin, hur,hur-review} for the relationship
of the von Neumann entanglement entropy with the QPT in the SBM, no
cusp-like behavior happens in our work for the entanglement changing
with respect to $\alpha''$ and our results could be applied to all
types of the bath. This is understandable because what we demonstrate is the scaling behavior
around the critical points for the parity breaking, instead of the
QPT for localization. Nevertheless, similar to the results in \cite
{chin,hur-review}, we also find that the maximal entanglement
appears when approaching the point $\alpha=\alpha_{c}$ from the
delocalization side and then a rapid disentanglement at the
localization side.

In summary, we have indicated the parity breaking in the SBM and
investigated analytically the scaling behavior of the magnetization
and the entanglement as well as  their relationship in the
neighborhood of the critical points for the parity breaking after
suppressing the intrinsic infrared divergence in the spectral
function. We argue that the conventionally employed spectral function is
not fully reasonable and the previous conclusions drawn for the QPT
happening in the Ohmic and sub-Ohmic SBM need more serious
reexamination. Our analytical treatment for the scaling behavior is
suitable for all types of the bath and  should be of general
interest, which is helpful for clarifying different numerical
results in previous publications and for understanding the phenomena
due to parity breaking and the physics hidden by the infrared
divergence.

This work is supported by funding from WIPM, by National Fundamental
Research Program of China (Grant No. 2012CB922102), and by NNSFC
under Grants No. 10974225 and No. 11004226.

\section*{Supplementary Material}

\section{Analytical solution to spin-boson model}

The SBM hamiltonian is given by \cite{leggett}
\begin{equation}
H=\frac {\epsilon}{2}\sigma_{z}-\frac {\Delta}{2}\sigma_{x}+
\sum_{k}\omega_{k}a^{\dagger}_{k}a_{k} + \sum_{k}\lambda_{k}(a^{\dagger}_{k}
+ a_{k})\sigma_{z},
\end{equation}
where $\sigma_{z}$ and $\sigma_{x}$ are usual Pauli operators,
$\epsilon$ and $\Delta$ are, respectively, the local field
(also called c-number bias \cite {leggett}) and
tunneling regarding the two levels of the spin. $a^{\dagger}_{k}$
and $a_{k}$ are creation and annihilation operators of the bath
modes with frequencies $\omega_{k}$, and $\lambda_{k}$ is the
coupling between the spin and the bath modes, which is governed by
the spectral function $J(\omega)=\pi\sum_{k}\lambda_{k}^{2}\delta
(\omega-\omega_{k})$ for $0<\omega <\omega_{c}$ with the cutoff
energy $\omega_{c}$.

We suppose the eigenfunction of Eq. (9) to be
$$|\Psi\rangle=
\begin{pmatrix}
\sum_{\{n\}} c_{\{n\}} |\{n\}\rangle_{A} \\
\sum_{\{n\}}(-1)^{\sum_{k}n_{k}+1}d_{\{n\}}|\{n\}\rangle_{B}
\end{pmatrix},$$
where $c_{\{n\}}$ and $d_{\{n\}}$ are coefficients to be determined
later and $\{n\}=n_{1}, \cdots, n_{N}$ are for different Bosonic
modes. $|\{n\}\rangle_{A(B)}$ is the product of displaced coherent
states of different modes \cite {Zhang}, i.e.,
$|\{n\}\rangle_{A(B)}=\prod_{k=1}^{N}|n_{k}\rangle_{A_{k}(B_{k})}$,
where
\begin{eqnarray}
|n_{k}\rangle_{A_{k}} =\frac {e^{-q_{k}^{2}/2}}{\sqrt{n_{k}!}}(a^{\dagger}_{k}+q_{k})^{n_{k}}e^{-q_{k}a_{k}^{\dagger}}|0\rangle, \notag\\
|n_{k}\rangle_{B_{k}} =\frac {e^{-q_{k}^{2}/2}}{\sqrt{n_{k}!}}(a^{\dagger}_{k}-q_{k})^{n_{k}}e^{q_{k}a_{k}^{\dagger}}|0\rangle, \notag
\end{eqnarray}
with the displacement variables $q_{k}=\lambda_{k}/\omega_{k}$ and
$k=1, 2, \cdots, N$. Using Schr{\"o}dinger equation, we have
\begin{equation}
\Big[\sum_{k}\omega _{k}(m_{k}-q_{k}^{2})+\frac {\epsilon}{2}\Big]c_{\{m\}} + \frac{\Delta }{2} \sum_{\{n\}}
d_{\{n\}}D_{\{m,n\}} =Ec_{\{m\}},
\end{equation}
\begin{equation}
\Big[\sum_{k}\omega _{k}(m_{k}-q_{k}^{2})-\frac {\epsilon}{2}\Big]d_{\{m\}} + \frac{\Delta }{2} \sum_{\{n\}}
c_{\{n\}}D_{\{m,n\}} =Ed_{\{m\}},
\end{equation}
where $D_{\{m,n\}}$ is given by \cite {Zhang,TLiu}
$$e^{-2\sum_{k}q_{k}^{2}}\prod_{k=1}^{N}\sum_{j=0}^{\min
[m_k,n_k]}(-1)^{j}\frac{\sqrt{m_{k}!n_{k}!}(2q_{k})^{m_{k}+n_{k}-2j}}{(m_{k}-j)!(n_{k}-j)!j!}.$$

Eqs. (10) and (11) are in principle solvable, but time- and resource-consuming using currently
available computing technology. For our purpose, under the condition $\Delta \ll$1, the terms of
$D_{\{m,n\}}$ with $\{m\} \neq \{n\}$ play negligible roles in the equations compared to other
terms with $\{m\} = \{n\}$ (The validity of the negligence of those terms is tested numerically below
in Figs. 6 and 7). So Eqs. (10) and (11) can be reduced to
\begin{equation}
\Big[\sum_{k}\omega _{k}(m_{k}-q_{k}^{2})+\frac {\epsilon}{2}\Big]c_{\{m\}} + \frac{\Delta }{2}
d_{\{m\}}D_{\{m,m\}} =Ec_{\{m\}},
\end{equation}
\begin{equation}
\Big[\sum_{k}\omega _{k}(m_{k}-q_{k}^{2})-\frac {\epsilon}{2}\Big]d_{\{m\}} + \frac{\Delta }{2}
c_{\{m\}}D_{\{m,m\}} =Ed_{\{m\}},
\end{equation}
from which we may straightforwardly obtain the analytical expressions of the eigenenergies
$E_{\{m\}}^{\pm}=\sum_{k}\omega_{k}(m_{k}-q_{k}^{2})\pm\sqrt{\epsilon^{2}+\Delta^{2}D^{2}_{\{m,m\}}}/2,$
and the coefficients
$c^{\pm}_{\{m\}}=\mu^{\pm}_{\{m\}}/\sqrt{1+(\mu^{\pm}_{\{m\}})^{2}}$
and $d^{\pm}_{\{m\}}=1/\sqrt{1+(\mu^{\pm}_{\{m\}})^{2}}$ with
$\mu^{\pm}_{\{m\}}=\Big[\epsilon\pm\sqrt{\epsilon^{2}+\Delta^{2}D^{2}_{\{m,m\}}}\Big]/\Delta
D_{\{m,m\}}$.

\section{Validity of the truncation for the bosonic modes }

In the latter half of the manuscript, we calculate the scaling behavior of the magnetization and the
entanglement by only considering the ground-state of the bosonic field, i.e., $\sum n_{k}=0$. To check if this truncation works well in the case of small
$\Delta$, we have calculated the magnetization with the bath modes of the discretized spectrum, based on the NRG
logarithmic discretization \cite {Bu1}, where we used our modified spectral function and compared different
truncations of the bosonic modes. Figs. 1 and 2 present that the magnetization remains unchanged under different truncation of the bosonic modes for different bath types, which indicate our calculation using only the ground-state of the bosonic field to be in saturation for the problem. So we may consider the results based on our analytical treatment with $\sum n_{k}=0$ to be reliable in the case of very small $\Delta$. Moreover, in the calculations for Figs. 1 and 2, we employed Eqs. (12) and (13) in the case of $\sum n_{k}=0$, and Eqs. (10) and (11) for the case of $\sum n_{k}=$1, 2 and 3. So the good fitting of the curves in the figures is also the justification of the approximation made in Eqs. (12) and (13).

\begin{figure}[!htpb]
\label{Fig1}
\includegraphics[width=3 in]{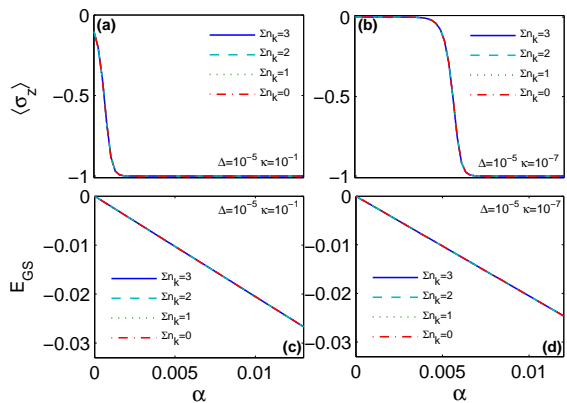}
\caption{The unchanged magnetization with respect to $\alpha$ under sub-Ohmic dissipation $s=0.2$ for different truncation of the bosonic modes, where
$\Lambda=2$, $\omega_{c}=1$ and we have considered the total bosons to be 0, 1, 2 and 3, respectively. }
\end{figure}
\begin{figure}[!htpb]
\label{Fig2}
\includegraphics[width=3 in]{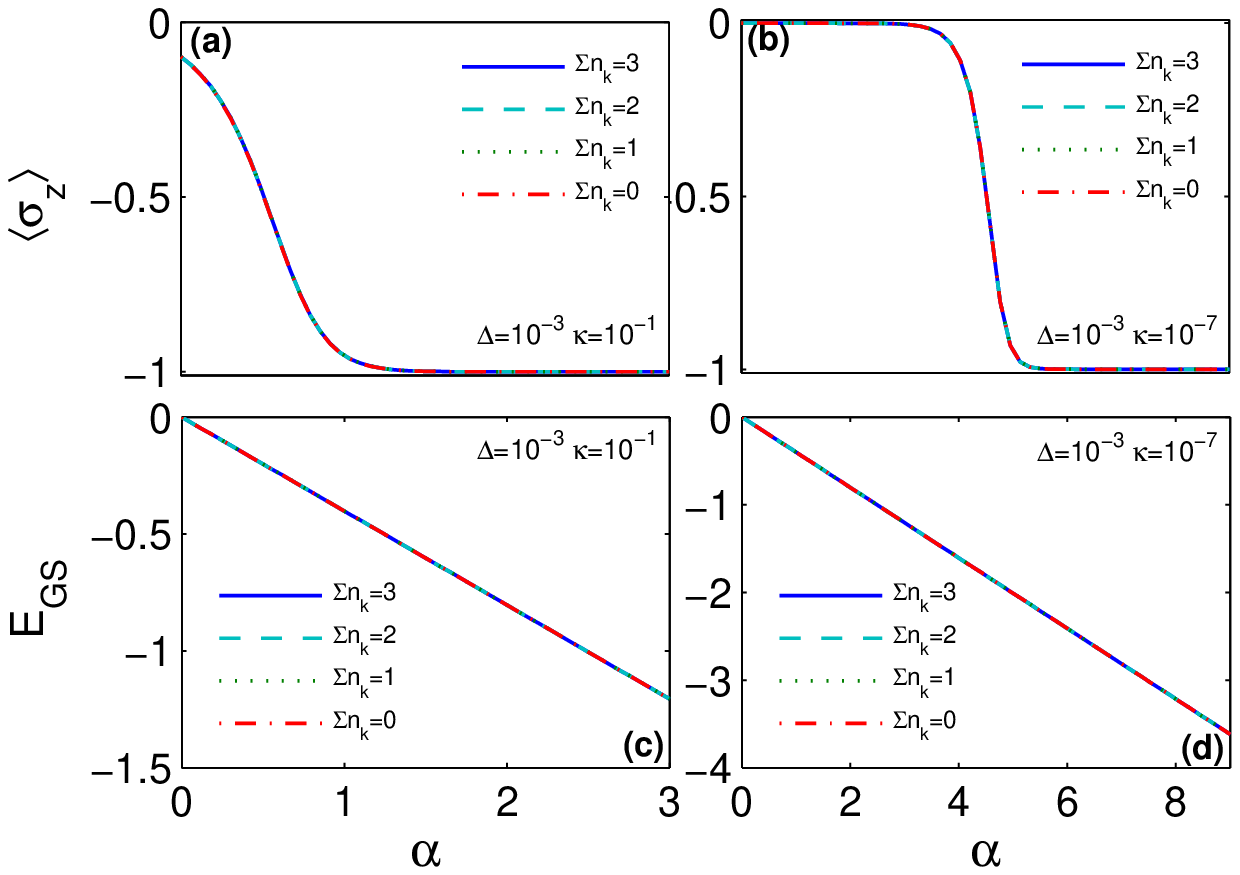}
\caption{The unchanged magnetization with respect to $\alpha$ under super-Ohmic dissipation $s=1.2$ for different truncation of the bosonic modes, where
$\Lambda=2$, $\omega_{c}=1$ and we have considered the total bosons to be 0, 1, 2 and 3, respectively. }
\end{figure}
\begin{figure}[!htpb]
\label{Fig3}
\includegraphics[width=3.0 in]{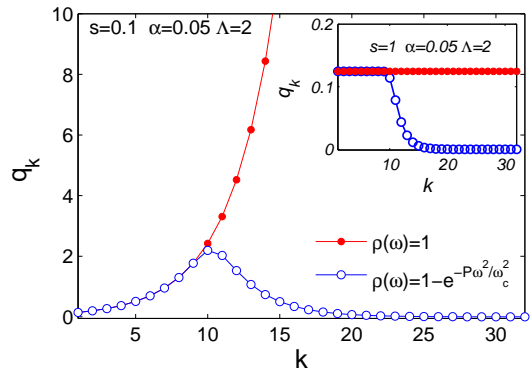}
\caption{(color online) $q_{k}$ as the function of $k$ under the treatment of the NRG logarithmic discretization using
$J(\omega)$ (the curve with red dots) and $J'(\omega)$ (the curve with blue circles), where $s=0.1$,
$\alpha=0.05$, $P=10^{6}$, $\omega_{c}=1$ and $\Lambda=2$. The inset is for $s=1$ with the same values of $\alpha$, $P$
and $\Lambda$. }
\end{figure}

\section{Calculation for bath modes of the discretized spectrum}

For the bath modes of the discretized spectrum, we modify the parameters in Eq. (9) by the NRG
logarithmic discretization \cite {Bu1} as
$$\omega_{k}=\xi_{k}=\gamma_{k}^{-2}\int_{\Lambda^{-(k+1)}\omega_{c}}^{\Lambda^{-k}\omega_{c}}\omega
J(\omega)d\omega$$ and $\lambda_{k}=\gamma_{k}/2\sqrt{\pi}$ with
$\gamma_{k}^{2}=\int_{\Lambda^{-(k+1)}\omega_{c}}^{\Lambda^{-k}\omega_{c}}J(\omega)d\omega$.
In such a case, it is easy to find $\sum_{k}q_{k}^{2}=\sum_{k}\gamma_{k}^{2}/(4\pi\xi_{k}^{2}),$ which, as
demonstrated in Fig. 8 for $s\le 1$, is divergent with $J(\omega)$ but convergent using $J'(\omega)$.
So our modified spectral function could effectively suppress the infrared divergence in the study of the SBM.


\begin{thebibliography}{99}
\bibitem{explain0} The infrared divergence also exists in other terms of $D_{\{m,n\}}$ with $\{m\},\{n\} >$0.
In our following analytical treatment, however, we only consider the ground-state bosonic mode, i.e., $\sum n_{k}=0$,
which could present an identical description, in the case of small tunneling, to the situation involving excited-state
bosonic modes, as shown in Sec II in Suplementary Material.
\bibitem{explain2} Despite the similarity between the cases of $\epsilon=0$ and $\Delta=0$,
we could not simply use the expression
$\alpha_{c}=-\ln(2\kappa^{2})/8\beta'$ for the case of $\Delta=0$,
which yields $\alpha_{c}\rightarrow\infty$. The correct way to
understand the situation of $\Delta=0$ should return to Eqs. (3) and
(4), and we could easily obtain the trivial solution for $\Delta=0$,
i.e., constant localization of the spin.
\end{thebibliography}
\end{document}